\documentclass[aps,amsfonts,amssymb,amsmath,preprint,longbibliography,prb,12pt,noshowpacs,]{revtex4-1}
\usepackage{amsfonts, amsmath, amssymb}
\usepackage{graphicx, xcolor}
\usepackage{stix}%

\begin{document}
	
 \title{Accelerate Microstructure Evolution Simulation Using Graph Neural Networks with Adaptive Spatiotemporal Resolution}
	\author{Shaoxun Fan}
    \author{Andrew L. Hitt}
	\author{Ming Tang}
	\email{mt20@rice.edu}
	\affiliation{Department of Materials Science and NanoEngineering, Rice University, Houston, TX 77005, USA}
	\author{Babak Sadigh}	
	\author{Fei Zhou} 
	\email{zhou6@llnl.gov}
	\affiliation{Physical and Life Sciences Directorate, Lawrence Livermore National Laboratory, Livermore, CA 94550, USA} 
	
	
	\newcommand{\mtcomment}[1]{\textcolor{magenta} {[MT: #1]}}
	\newcommand{\kycomment}[1]{\textcolor{cyan} {[KY: #1]}}
	\newcommand{\mtrevision}[1]{\textcolor{black} {#1}}
	\newcommand{\finalrevision}[1]{\textcolor{blue}{#1}}
	
	\begin{abstract}
	Surrogate models driven by sizeable datasets and scientific machine-learning methods have emerged as an attractive microstructure simulation tool with the potential to deliver predictive microstructure evolution dynamics with huge savings in computational costs. Taking 2D and 3D grain growth simulations as an example, we present a completely overhauled computational framework based on graph neural networks with not only excellent agreement to both the ground truth phase-field methods and theoretical predictions, but enhanced accuracy and efficiency compared to previous works based on convolutional neural networks. These improvements can be attributed to the graph representation, both improved predictive power and a more flexible data structure amenable to adaptive mesh refinement. As the simulated microstructures coarsen, our method can adaptively adopt remeshed grids and larger timesteps to achieve further speedup.
 The data-to-model pipeline with training procedures together with the source codes are provided.
	\end{abstract}
	\maketitle

\section*{Introduction}
Macroscopic properties of materials are not solely determined by their atomic structures but also by their microstructures, or the arrangement of material constituents, such as grains and phases, at the mesoscopic level. A key achievement of modern materials science is the tailoring of materials properties by modifying microstructure through various processing techniques such as casting, annealing and forging to meet specific requirements. To this end, the ability to predict how microstructure evolves dynamically under different processing conditions is crucial to the rational design and development of advanced materials.

Simulating microstructure evolution typically relies on continuum models that employ partial differential equations (PDE) to describe the evolution rules of the systems.
For instance, mesoscale modeling methods such as phase-field models (PFM) are widely applied in the studies of solidification, solid phase transformations, grain growth, etc.~\cite{sagui1998,Collins1985,Steinbach1996,Nishimori1990}
However, such simulations are computationally expensive to perform even with the use of the state-of-the-art computer hardware. This restricts the time and length scales that could be achieved in simulations and limits their usefulness in predicting the processing-structure-property relation at the macroscopic level. As such, extensive efforts have been devoted to the development of acceleration techniques for microstructure simulations such as parallel computing and adaptive mesh refinement (AMR)~\cite{Greenwood2018,Ceniceros2010,Guo2015}, but the gap remains significant.  Notably the Markov Random Fields (MRF) method, a powerful and scalable computational approach, has been recently developed for accurate 2/3D polycrystalline reconstruction and spatiotemporal grain-growth predictions based on experimental inputs \cite{Acar2016MSMSE, Javaheri2020CAD} and can be integrated into component-sized design workflow \cite{Javaheri2022CMS}.

Scientific machine learning (ML), including deep learning methods, has emerged in recent years as a powerful family of computational tools that complement and extend the capabilities of traditional computational materials science toolkit \cite{Butler2018N, Zhang2023-AI4Science}. In the domain of microstructure characterization, convolutional neural networks (CNN) and related methods have been successfully applied to various image analysis tasks such as feature (grain size, aspect ratio, spacing, etc) extraction and quantification, microstructure classification, image denoising and super-resolution, defect detection, semantic image segmentation, and even 2D and 3D microstructure generation \cite{Alrfou2022,Zhu2022,Henkes2022,Zhao2023,Gorynski2023,Jung2021,Shen2021,Khurjekar2023}. As such, CNN-based deep learning methods are becoming part of the standard toolkit for static image processing for microscopy experiments. Very recently, CNNs combined with recurrent neural networks (RNN) have been proposed to emulate the PDEs underlying PFM models to improve the computational efficiency of microstructure evolution simulations \cite{Yang2021P, MontesdeOcaZapiain2021nCM, Wu2023CMS, KazemzadehFarizhandi2023CMS}.
In Ref.~\cite{Yang2021P}, Yang et al.\ demonstrated that a convolutional RNN model (ConvRNN) based on the E3D LSTM architecture~\cite{Wang2019-E3D} is able to learn the underlying evolution rules from 2D PFM simulation data and quantitatively reproduce PFM predictions. The trained CNN model accurately captures short-term local dynamics and long-term statistical properties in several common microstructure evolution processes including grain growth, spinodal decomposition and dendrititic crystal growth.
Furthermore, the model exhibits excellent capability in extrapolating with robust and transferable accuracy to new configurations beyond the training datasets.
Compared to PFM, ConvRNN-based simulations are faster by 1 to 3 orders of magnitude, which derives from the use of larger time steps and coarser mesh grid without sacrificing the prediction accuracy.

This work aims to improve both the accuracy and efficiency over our previous CNN-based ML surrogate models \cite{Yang2021P} for microstructure evolution. This is achieved with an overhaul of the basic model architecture: instead of ConvRNN, we propose a graph neural network (GNN) model to fulfill the goal of performing longer and larger mesoscale simulations using grain growth as an example. Graph is a basic data structure that contains two components: vertices and edges. GNN is a type of emerging neural network that aims to process data in the form of graphs and became popular in the last decade \cite{battaglia2018relational, Sanchez-Gonzalez2020}. When it comes to the fields of chemistry and materials science, GNN is a natural choice for atomic level studies, as atoms and chemical bonds can be directly interpreted as vertices and edges, and variants of GNN were developed for predicting properties of crystals, molecules and polycrystalline materials \cite{Dong2023,Cheng2021, Dai2021}. A multitude of GNN-based interatomic potential models have emerged that can be trained from accurate but expensive quantum-mechanical calculations to predict quantum-accurate forces/energies and enable highly accurate molecular dynamics simulations \cite{Batzner2022,Chen2022,Li2022}.

Simulating complex physics with GNNs is becoming an attractive alternative to CNN-based surrogate models in mesh-based fluid simulations \cite{Sanchez-Gonzalez2020, Pfaff2020GNN}.
MeshGraphNet (MGN) \cite{Pfaff2020GNN}, a recent GNN variant, has proved well suited for representation of complex physics \cite{Bertin2023-DDGNN, Bertin2023-MLDDD}. In an MGN, the values of each vertex is passed to its nearest connected neighbors through some learned function in the message passing layer (MPL), with multiple MPLs creating a local spatial awareness. The graph-based architecture allows for more flexible geometric representations when compared to CNNs \cite{Yang2021P}. Very recently, Xue {\textit{et al}} adopted a graph-based representations to solve multi-physics PFM of microstructure evolution in additive manufacturing with 50x speedup \cite{Xue2022nCM}. Qin {\textit{et al}} developed a long-short-term-memory based surrogate model of 2D epitaxial grain growth in additive manufacturing conditions \cite{Qin2023CMS}.

As discussed previously \cite{Yang2021P}, the history-dependence of grain growth problems is weak and can be ignored without impact on accuracy. We also limit the scope of this paper to isotropic grain growth to make direct comparison to Ref.~\cite{Yang2021P}. In this work, we adopt a feedforward MGN for time series prediction without recurrency. Hereafter we will refer to our previous ConvRNN as simply CNN to emphasize the main difference in model architecture. The rest of the paper is organized as follows. The main results on the accuracy of the GNN model are presented and compared to the baseline CNN model for 2D and 3D, respectively. We then make a deep dive into the computational efficiency, including direct speed comparison, and strategies to accelerate GNN-based simulation in the spatial and temporal dimensions. More specifically, we present design and use of a simple adaptive mesh refinement algorithm enabled by the flexible graph data structure, and an adaptive time stepping algorithm suitable for coarser grains in the later stages of growth. The computational details are presented in the methods section.

\section*{Results}
\subsection*{Grain growth in 2D: GNN vs CNN}
Grain growth (GG) is a classic example of microstructure evolution during high-temperature processing of polycrystals.
It refers to the phenomenon of the average grain size continuously increasing with time due to the growth of some grains at the expense of others.
Grain boundary (GB) movement during GG is driven by the reduction of GB energy,
and the local GB velocity could be expressed as   
\begin{equation}
	v_{\text{gb}} = M\gamma\kappa
\end{equation}
where $M$ is the GB mobility, $\gamma$ is the GB energy and $\kappa$ is the local boundary curvature.
For GG in 2D systems, Smith and von Neumann ~\cite{von1952metal} and Mullins~\cite{mullins1956two} showed that the rate of area change of an individual grain obeys a remarkably simple ``N-6'' rule:
\begin{equation}
	\frac{dA}{dt} = M\gamma\frac{\pi}{3}(N-6)
\end{equation}
where $N$ is the number of neighboring grains sharing a GB.
In 2D the growth rate depends only on topology and not on grain size.
A grain grows when it has more than six edges ($N>6$) or shrinks if $N<6$. 
Another important aspect of grain growth is that it could attain statistical self-similarity,
i.e. the normalized grain size distribution and grain shape distribution remain constant during growth.

Grain growth has been simulated with various approaches such as the phase-field method~\cite{Krill2002,Kim2006PRE-Computer,Moelans2008C,Moelans2009}, Monte-Carlo Potts model~\cite{Zollner2006SM,Ivasishin2006MSEA,Chang2012},
surface evolver~\cite{Marthinsen1996,Wakai2000AM}, etc.
Previously, we demonstrate that CNN could learn from phase-field simulation data to predict GG in 2D~\cite{Yang2021P}. 
Here we apply MGN to grain growth prediction and compare its performance with CNN.
The mesh GNN model consists of 11 message passing blocks, each of which contains 128 latent features.
The training set was the same as in our previous work \cite{Yang2021P}. 
Training data are obtained from 1520 phase-field simulations of isotropic 2D grain growth. 
Each simulation was performed on a 256$\times$256 grid with periodic boundary conditions with a different initial grain structure,
which contains about 75 grains. 
A 20-frame trajectory was obtained from a simulation, in which the number of grains decreases to $\sim$45. 
Each frame is a grayscale image (white: grain interior; black: grain boundary) that is downsampled from the phase-field result from 256$\times$256 to 64$\times$64 pixels by averaging. 
The time interval between two adjacent frames corresponded to 80 PDE time steps.
Another 1000 simulations are used as the test dataset. 

\begin{figure}[thp]
			\includegraphics[width=0.7\columnwidth]{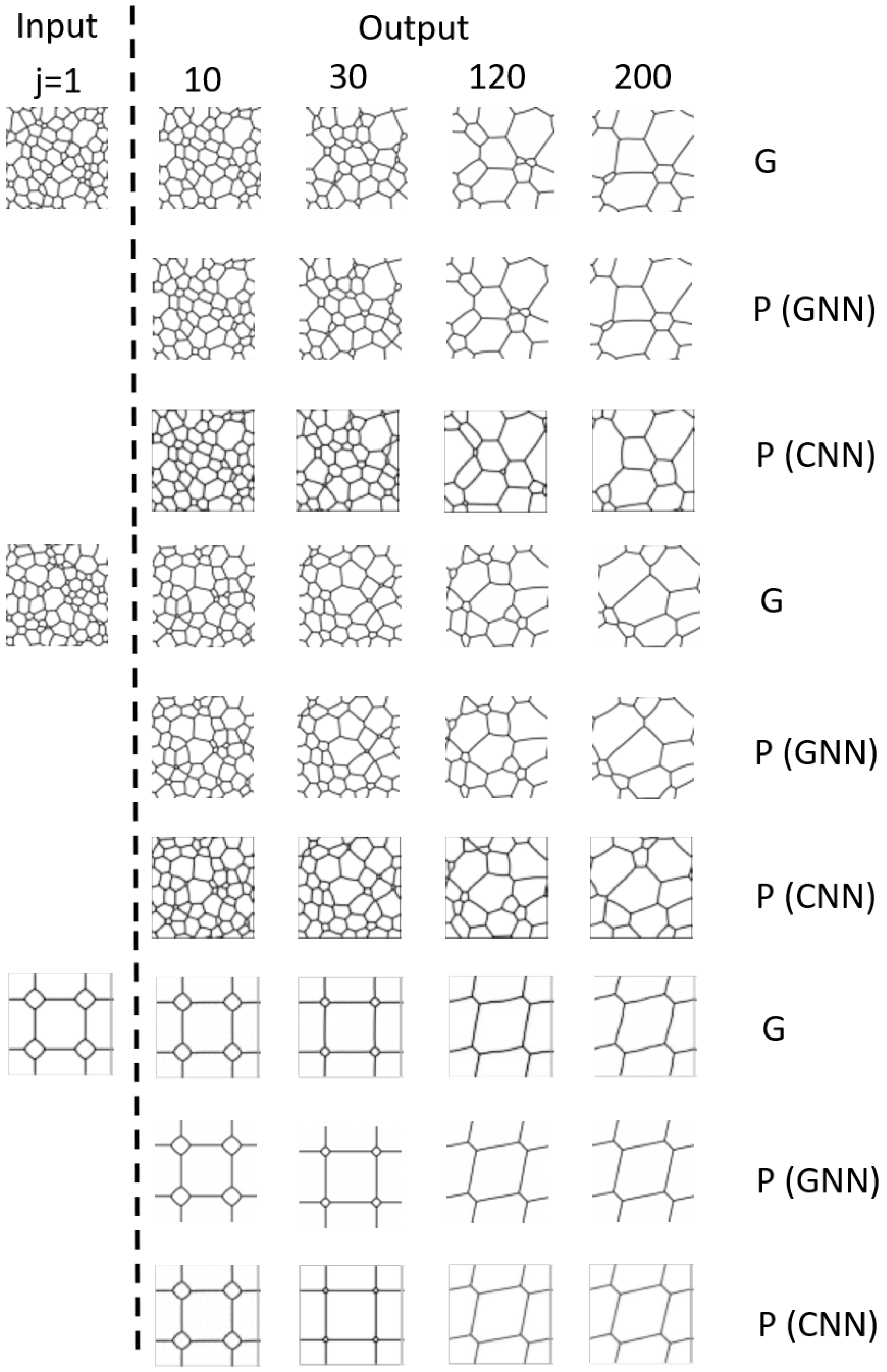}
			\caption{Examples of GNN prediction (P) based on 1 input frame in comparison with the ground truth (G) and previous CNN predictions \cite{Yang2021P}.}
			\label{fig:2D_prediction}
\end{figure}

\begin{figure}[thp]
	\includegraphics[width=0.9\columnwidth]{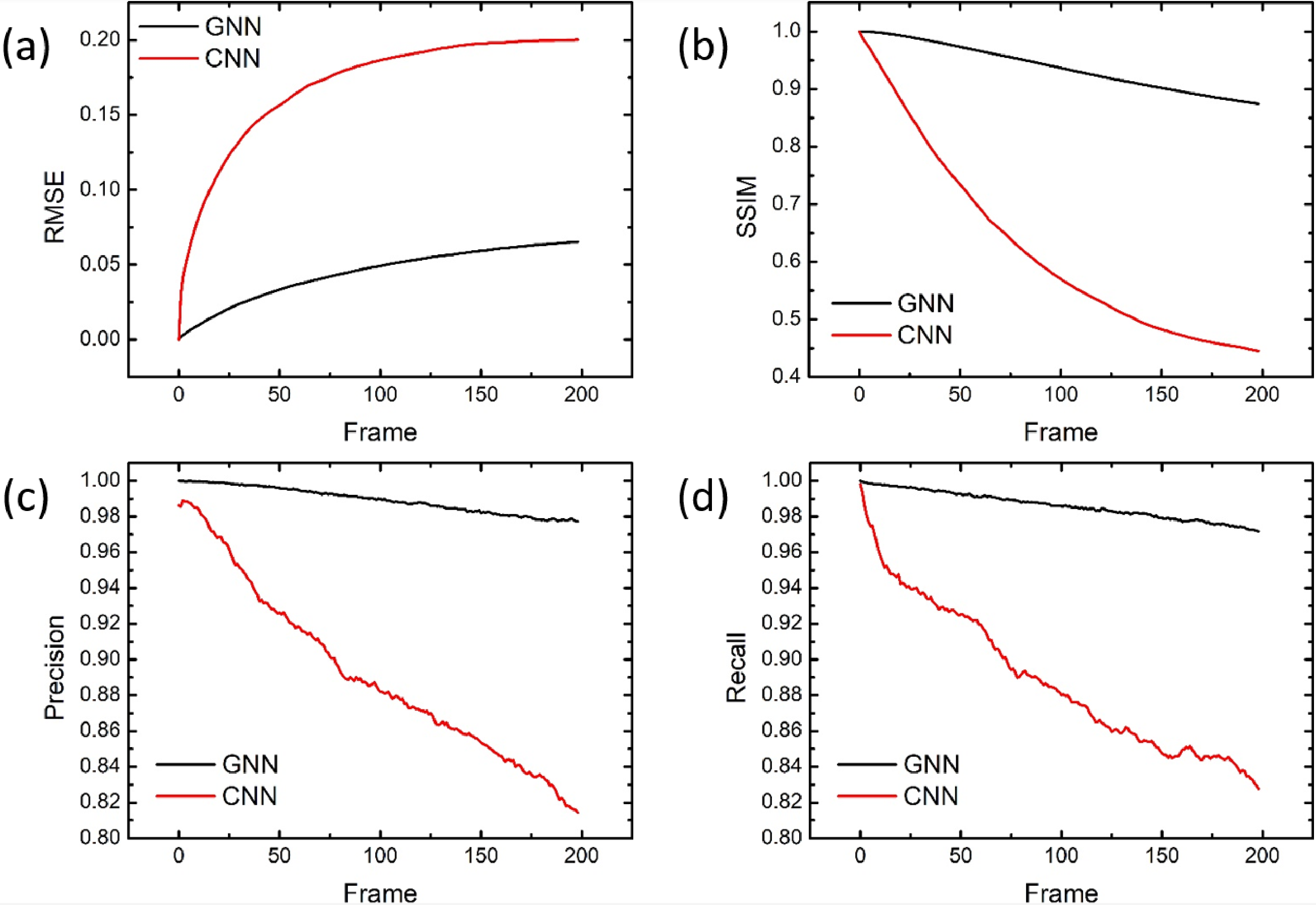}
	\caption{Comparison of the prediction accuracy of GNN vs CNN models for 2D grain growth: (a) RMSE, (b) SSIM, (c) Precision and (d) Recall. See main text for definitions of precision and recall.}
	\label{fig:2D_accuracy}
\end{figure}

As our focus is on the long-time prediction accuracy of ML, 
the trained GNN model was tasked to predict 200 frames of grain structure evolution based on different initial configurations. 
The predicted trajectories are thus 10 times longer than the training data,
i.e. 90 precent of the polycrystalline states were never seen by GNN during training.
Figure \ref{fig:2D_prediction} shows three examples of the GNN predictions along with the ground truth from phase-field simulations. In addition, the CNN model studied in our previous work was trained with the same data and its prediction is also compared against GNN.
The first two prediction examples starts from initial polycrystalline configurations that were generated by Voronoi tessellation. 
In these trajectories, while visible difference appears between the CNN prediction and ground truth at about the 30th frame, grain structure generated by GNN is visually indistinguishable from the ground truth even at the last (200th) frame.
In the third example, the GNN model was tested by inputing an artificial configuration which consists of four small 4-sided grains and four larger 8-sided grains. While the input configuration is different from grain structures in the training data, its evolution was still predicted accurately.
Figure \ref{fig:2D_accuracy}a and b compare the RMSE and SSIM (as explained in section Computional method F) averaged over the 1000 test cases predicted by GNN and CNN.
RMSE of GNN predictions remains below 0.07 after 200 frames, which is significantly smaller than CNN's error ($\sim$0.2).
Similarly, GNN achieves a much higher SSIM than CNN (0.87 vs. 0.45) after 200 frames. 
In addition to pixel-wise measurements, we evaluated the model accuracy using two classification-related metrics, precision and recall, which quantify how well the models perform in terms of predicting which grains survive or disappear. 
To this end, individual grains were labeled and tracked through an image sequence to determine whether a specific grain in the initial configuration remains in the system in a given frame.  
Precision is defined as $\text{TP}/(\text{TP}+\text{FP})$, and recall is $\text{TP}/(\text{TP}+\text{FN})$, where T(F)P(N) designates counts of true (false) positive (negative) classifications.
For 1000 GNN test cases, the average precision and recall start at 1 and slowly decrease to 0.97-0.98 at the end of the trajectories as shown in Figure \ref{fig:2D_accuracy}c and d. 
Given that there are approximately 10 grains left in the last (200th) frame, 
this suggests that GNN correctly predicts all the surviving grains in the  vast majority of the trajectories.
In contrast, CNN incorrectly forecasts ~20\% of the grains in the final grain structure.  
The above comparisons convincingly establish that GNN outperforms CNN in grain growth predictions and has significantly better long-term accuracy.

\begin{figure}[thp]
	\includegraphics[width=0.9\columnwidth]{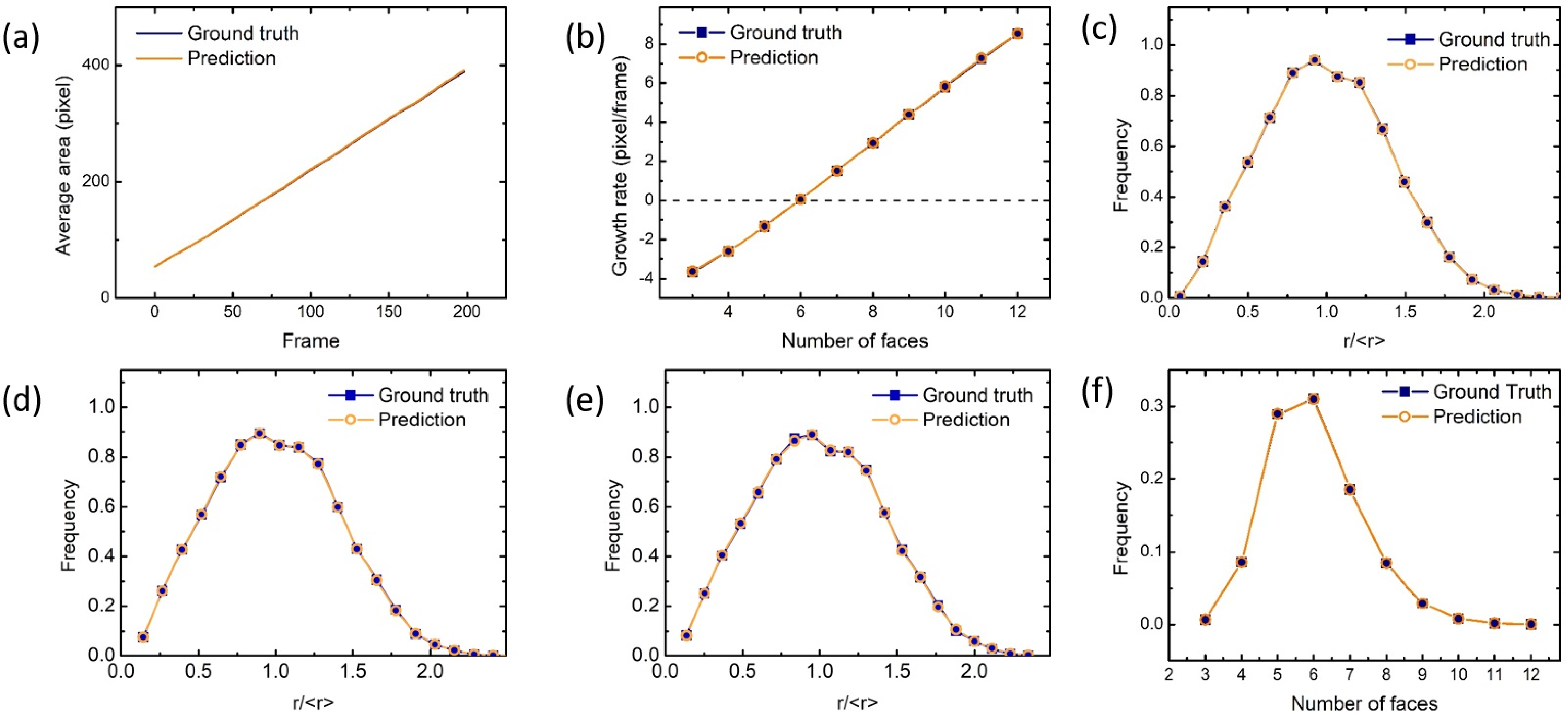}
	\caption{Statistical properties of 2D grain growth predicted by GNN compared with ground truth. (a) Average grain area $A$ as a function of time; (b) Dependence of the average grain growth rate $dA/dt$ on the number of grain faces. (c-f) Steady-state distribution of normalized grain size $r/<r>$ ($r = \sqrt{A/\pi}$) computed from (c) all the frames, (d) 50th frame, (e) 100th frame and (f) number of grain faces computed from all the frames.}
	\label{fig:2D_stats}
\end{figure}

Beyond pixel- and grain-based measures, GNN also precisely captures the statistical features of the grain growth process.
It can be seen from Figure \ref{fig:2D_stats} that the GNN predictions of average grain area vs time (a), 
grain growth rate vs number of grain sides (b), grain size distribution averaged over all the frames (c) or at different time steps (d,e) and number of grain side distribution (f)
are almost indistinguishable from the ground truth. 
The observed agreement is again superior to that of CNN \cite{Yang2021P}.

\begin{figure}[thp]
	\includegraphics[width=0.9\columnwidth]{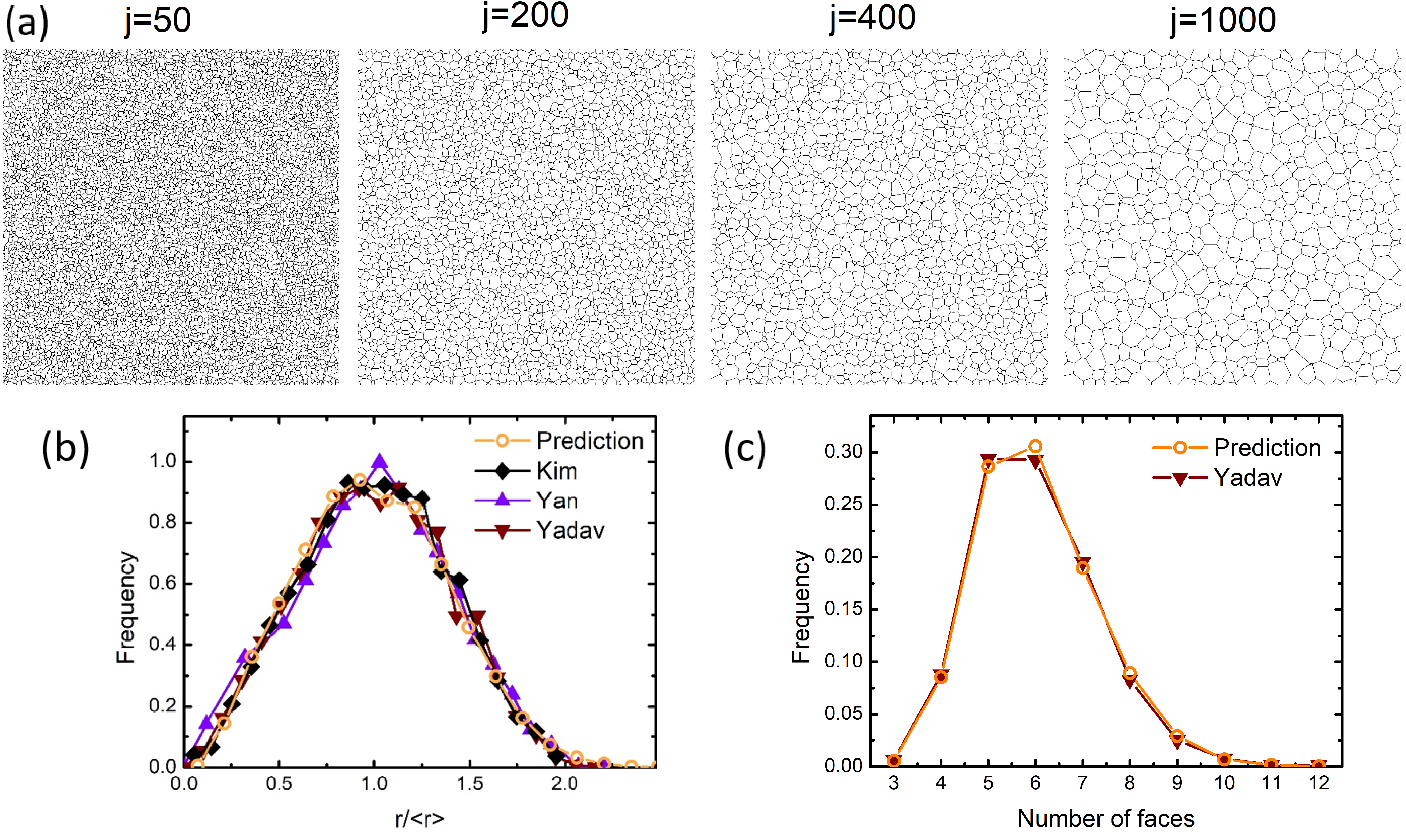}
	\caption{(a) 2D grain structure on a 1024$^2$ mesh at different times from GNN simulation; (b,c) Steady-state distribution of normalized grain size (b) and number of faces (c) predicted by GNN in comparison with phase-field simulation results in literature\cite{Kim2006PRE-Computer,Yan2022,Yadav2018}}
	\label{fig:1024}
\end{figure}

The GNN architecture is naturally extendable over space, e.g.\ a GNN model trained on 64$\times$64 grids could be applied to 1024$\times$1024 without retraining.
The spatiotemporal extrapolation capability of GNN means that it could be efficiently trained on a relatively small dataset and then used for large scale simulations that are computationally expensive for the phase-field method. 
As a demonstration, Figure \ref{fig:1024}a shows the evolution of a polycrystal structure of 1024$\times$1024 pixels from GNN predictions.  
The configuration is 256 times larger in area than the train data, and the prediction time span (1000 frames) is 50 times longer. 
The initial structure contains approximately 50000 grains.
We find that the grain size and number of sides distributions calculated from the GNN predictions (Figure~\ref{fig:1024}b and c) agree very well with results from phase-field simulations reported in literature~\cite{Kim2006PRE-Computer,Yan2022,Yadav2018}. 
Nevertheless, the computation cost of GNN is substantially lower than phase-field simulations, which will be compared in a later section.

\subsection*{Grain growth in 3D}
Next we apply GNN to learn and predict grain growth in 3D. 
An advantage of GNN is that the same architecture could be used for both 2D and 3D simulations since the input mesh data are first transformed into a graph regardless of their dimensionality. The baseline CNN model was generalized to 3D data by replacing 2D convolutions with 3D. %
Both GNN and CNN were trained with 300 trajectories of phase-field simulations performed on a 128$^3$ mesh and downsampled to 32$^3$ voxels.
Each trajectory contains 31 frames and has 350 grains in the initial configurations.
As the 3D requires more memory usage than the 2D counterpart, 
the GNN architecture is simplified to contain 10 message passing layers, each with 64 features, to reduce memory consumption.

\begin{figure}[thp]
	\includegraphics[width=0.7\columnwidth]{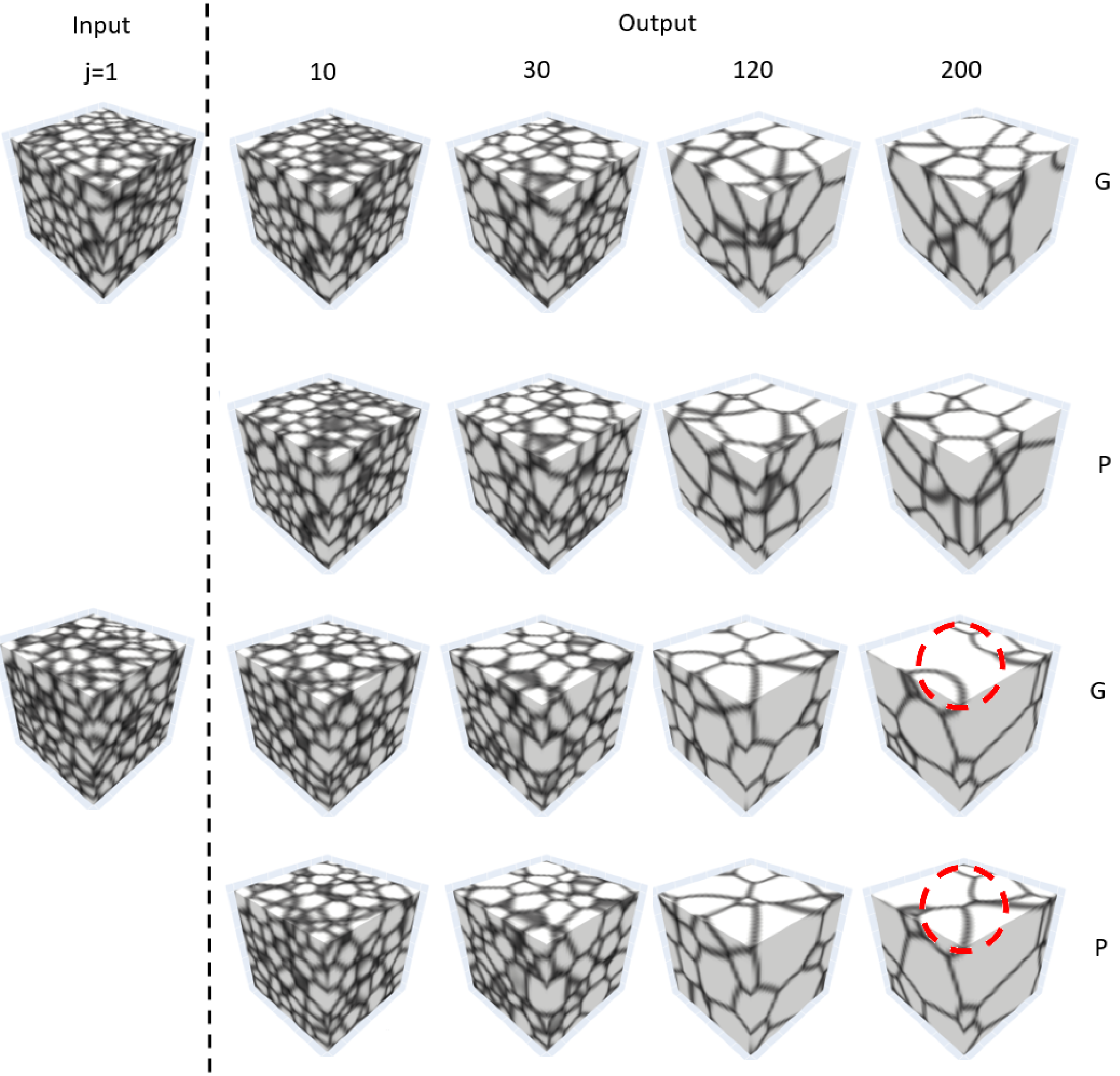}
	\caption{Examples of GNN prediction (P) of 3D grain growth on a $32^3$ mesh in comparison with the ground truth (G)}
	\label{fig:3D_prediction}
\end{figure}

\begin{figure}[thp]
	\includegraphics[width=0.9\columnwidth]{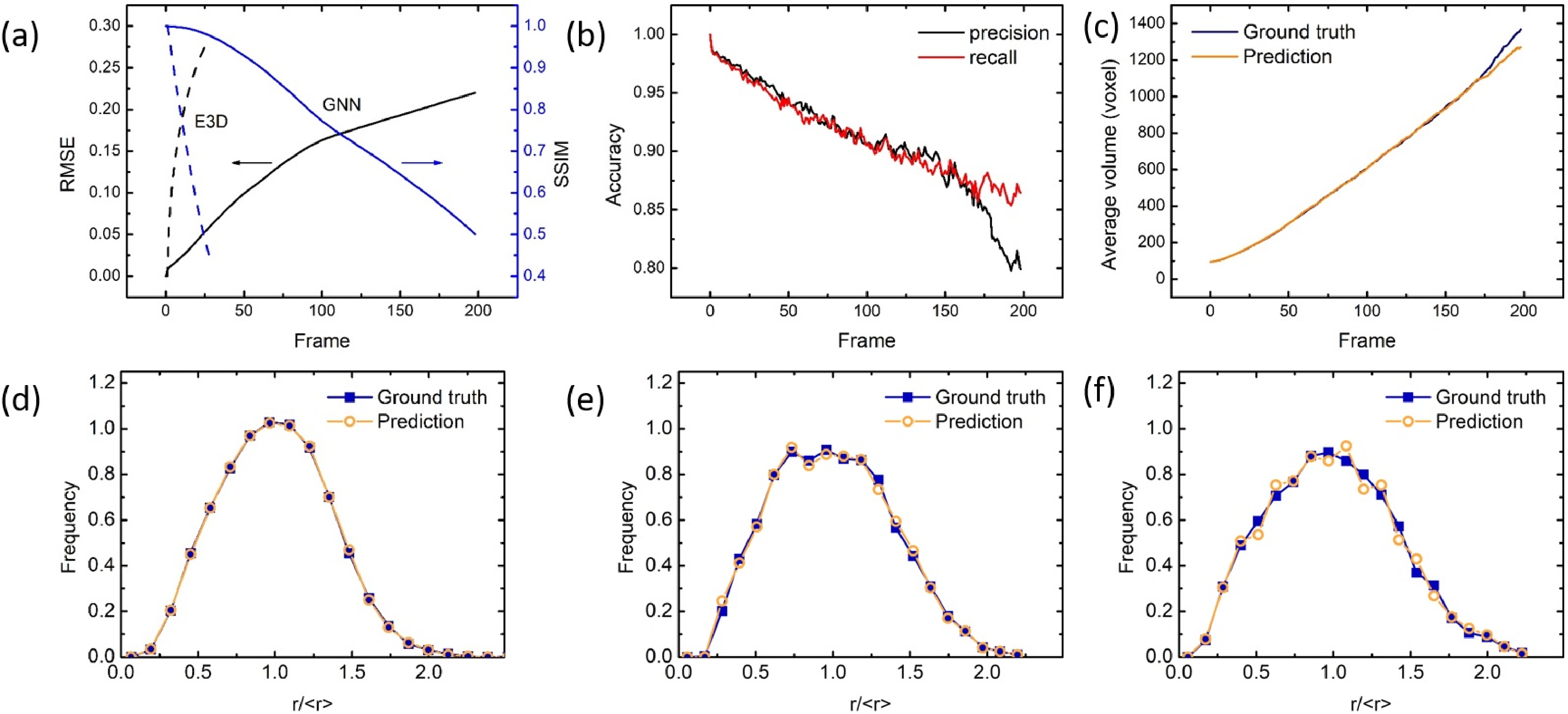}
	\caption{Accuracy of the GNN prediction of 3D grain growth: (a) RMSE and SSIM of predictions by GNN (solid lines) vs CNN (dashed lines); (b) Precision and recall of the surviving grains predicted by GNN; (c) Average grain area vs time predicted by GNN vs ground truth; Steady-state distribution of normalized grain size from the ground truth and GNN prediction computed from (d) all the frames, (e) 50th frame and (f) 100th frame.} 
	\label{fig:3D_analysis}
\end{figure}

After training,  the 3D model was tested to predict 200 frames. 
Two exemples of predicted trajectories with different initial grain configurations are compared with the ground truth in Figure \ref{fig:3D_prediction}. 
The predictions are found to bear close resemblance to the ground truth in the first 50 frames.
Though visible difference starts to emerge at the later stage, 
the predicted grain structures are free of any noticeable artifacts.
Figure~\ref{fig:3D_analysis}a shows that the average RMSE of 49 test trajectories increases to 0.22 at the 200th frame
while the average SSIM decreases to 0.50. 
However, the corresponding precision and recall metrics (Figure~\ref{fig:3D_analysis}b) remain higher than 85\% up to 170th frame, suggesting that the GNN correctly predicts the majority of surviving grains after long-time evolution. 
On the contrary, CNN fails to make 3D predictions, as shown by the dashed lines in Figure~\ref{fig:3D_analysis}a. Extensive attempts were made to adjust various hyperparameters of the 3D CNN model with none turning out to be predictive, even though comparable hyperparameters wherever applicable were adopted in 2D CNN and GNN training. The origin of the failure of CNN in 3D was not completely clear. Henceforth we will focus on the GNN results only.
The precision value exhibits an accelerated drop around 170th frame in Figure~\ref{fig:3D_analysis}b.
Such a seemingly abnormal behavior results from an artifact of the phase-field model and is not an indicator of GNN's poor performance. 
As illustrated by the grains highlighted by the red circle at $t$= 200 in Figure~\ref{fig:3D_prediction},
a grain could ``meet'' itself during growth in a PDE solver because of the periodic boundary condition   
when there are only a small number of grains left in the system.
In phase-field simulation, such an event causes the two regions of the same grain to merge into a contiguous domain, which was caused by the nature of periodic boundary condition used in simulations.
This grain self-coalescence , essentially a finite-size effect of PDE simulations when there are too few grains left, is  nonetheless absent in GNN predictions because the GNN model has no knowledge about grain identities and always assumes that any two separated regions are different. 
Instead, a grain boundary persists between the two regions as highlighted in Figure~\ref{fig:3D_prediction}. Such disagreement between GNN and ground truth PDE is therefore a finite-size artifact present only in the late stage of grain growth simulations with very few grains left.

Similar to 2D grain growth, the agreement between the 3D predictions and ground truth at the statistical level is excellent. 
Figure~\ref{fig:3D_analysis}c-f shows that the predicted average grain volume and grain size distribution are almost identical to the phase-field simulations with the exception of the aforementioned artifact after about 170 steps in Figure~\ref{fig:3D_analysis}c and slight discrepancy of the grain size distribution at 100th frame in Figure~\ref{fig:3D_analysis}f.
Compared to 2D predictions, GNN exhibits slightly lower long-term accuracy in predicting 3D grain growth. 
This could be attributed to a simpler neural net structure and a smaller training set used in 3D predictions,
which could be improved at the expense of more memory usage.    
 
\begin{figure}[thp]
	\includegraphics[width=1\columnwidth]{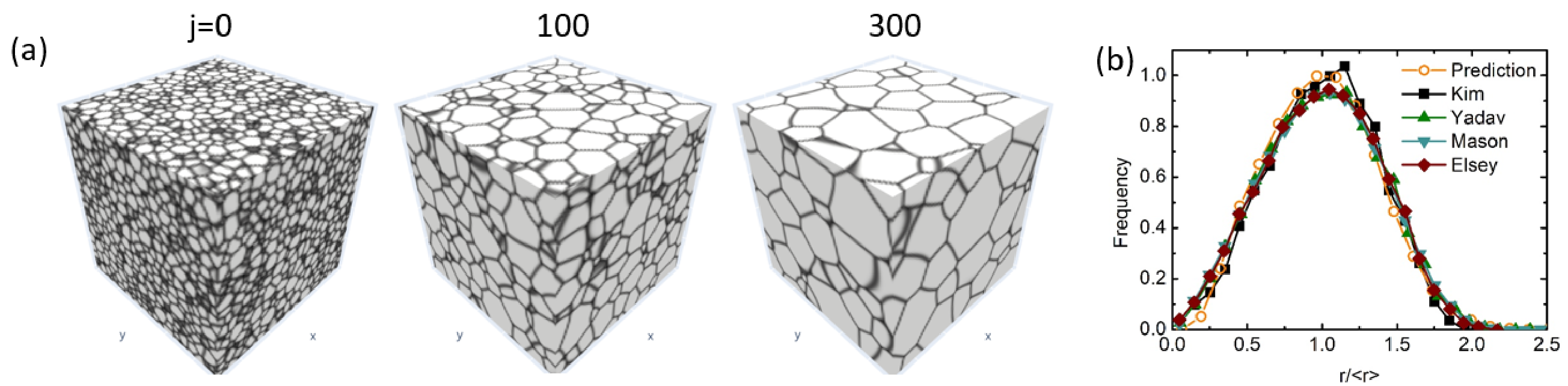}
	\caption{Predictions of the 3D GNN model on a large $96^3$ mesh: (a) trajectory; (b) The normalized steady-state grain size distribution calculated from the GNN predicted frames in comparison to results from phase-field simulations reported in literature\cite{Kim2006PRE-Computer,Yadav2018, Mason2015,Elsey2009}.}
	\label{fig:3D_largesimul}
\end{figure}

\begin{figure}[thp]
	\includegraphics[width=0.7\columnwidth]{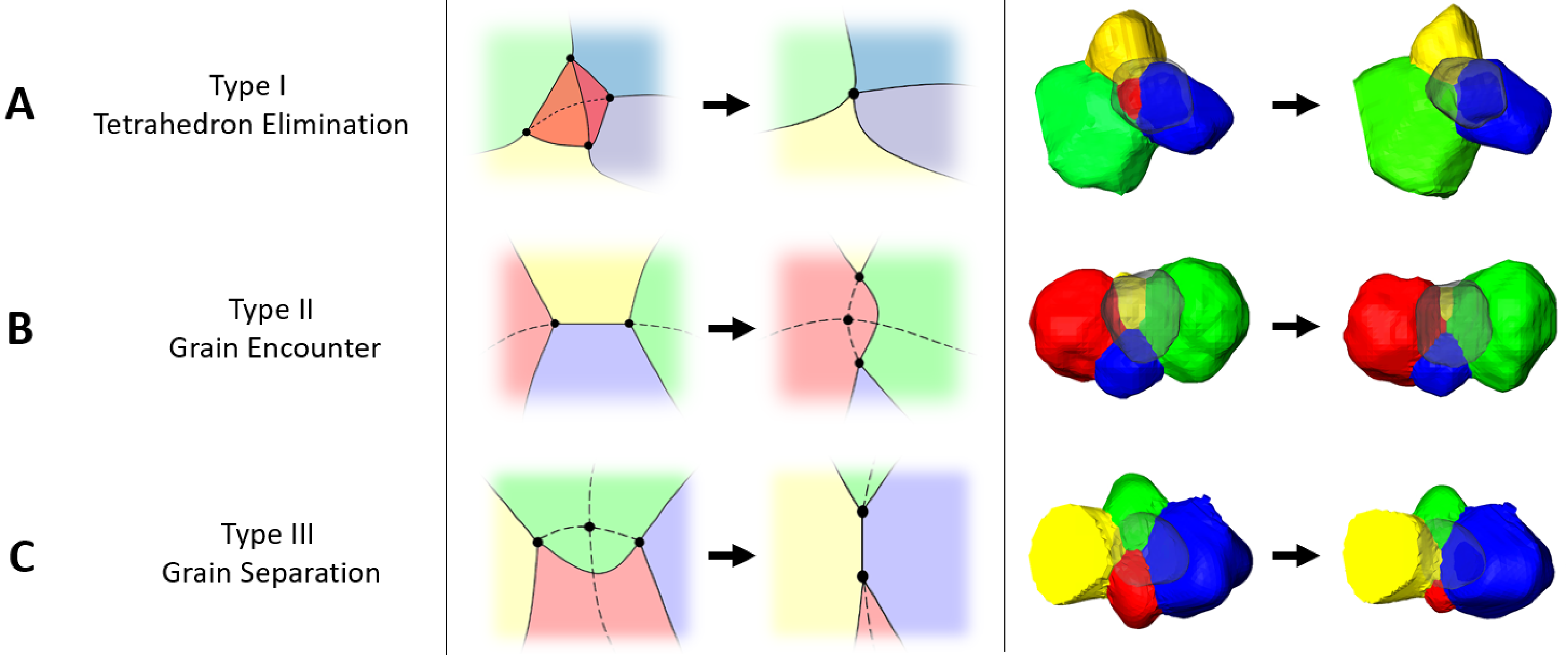}
	\caption{Three topological evolution events during grain growth in 3D (as described in \cite{Patterson2013AM-Schlegel}). The schematics of each event are shown in the second column (solid lines -- triple lines, dashed lines -- triple lines obscured by the front grains, and dots -- quadruple points). The third column shows the 3D rendering of an example of each event from GNN prediction in agreement with the ground truth.  
}
 \label{fig:3D_events}
\end{figure}

Large scale 3D simulations could be carried out using GNN with both precision and efficiency.    
As a demonstration, we apply the trained GNN model to a polycrystalline configuration containing 96$^3$ voxels, 
which corresponds to phase-field simulation on a 384$^3$ mesh due to downsampling. 
Snapshots of the grain structure at $t$= 0, 100, 300 are shown in Figure~\ref{fig:3D_largesimul}a-c. 
The steady-state grain size distribution is shown in Figure~\ref{fig:3D_largesimul}d, in excellent agreement with large-scale simulations of isotropic grain growth using the phase-field or front-tracking method~\cite{Kim2006PRE-Computer,Yadav2018, Mason2015,Elsey2009} as well as the mean-field theory prediction~\cite{Hillert1964}. 
Furthermore, there are three types of topological evolution events in 3D grain growth\cite{Patterson2013AM-Schlegel}: tetrahedral elimination, grain encounter and grain separation as illustrated in Figure \ref{fig:3D_events}. 
We found that all of these events are correctly captured by GNN in the trajectories.
We conclude that GNN's excellent predictive power derives from its ability to reliably predict the curvature-driven boundary movement and the 3D topological changes to the GB network.

\section*{Computation Efficiency}
\begin{figure}[thp]
 		\vskip 0.15in
 		\centering
		\includegraphics[width=0.7\columnwidth]{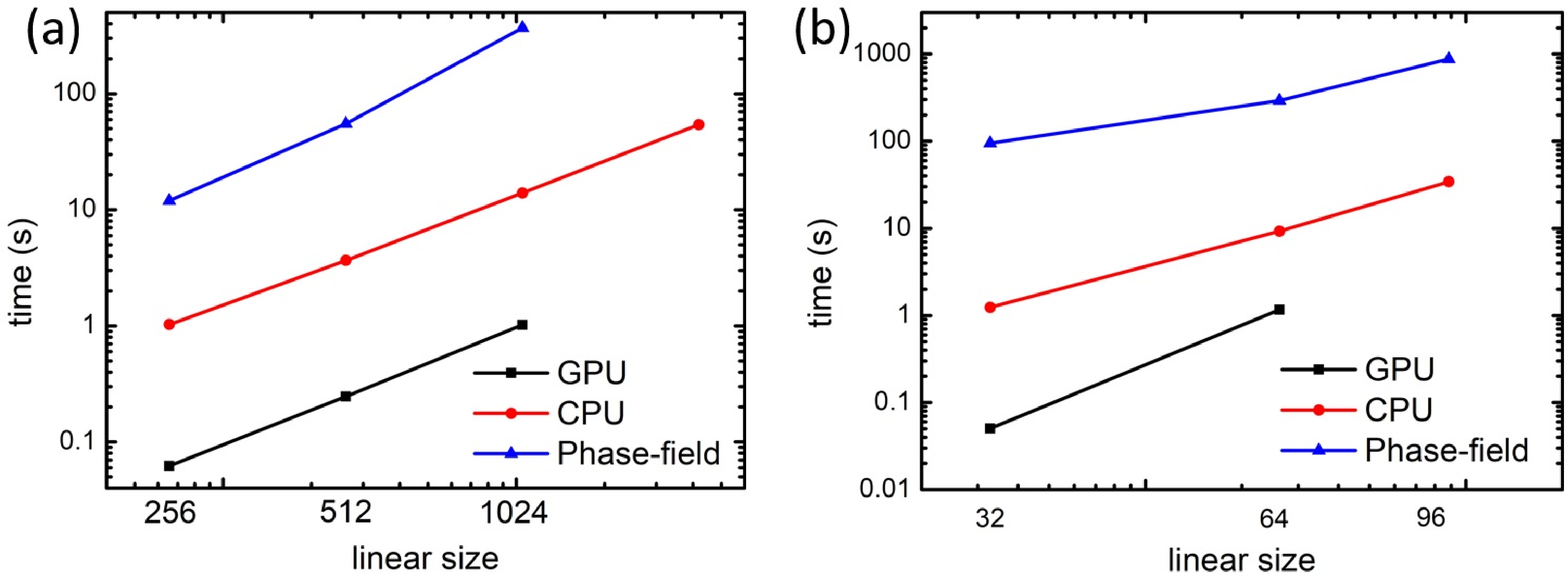}
		\caption{Average wall time per GNN step over 500 trials as a function of the computation domain size $L$ (downsampled) of (a) 2D and (b) 3D simulations.}
		\label{fig:scaling2D}
\end{figure}

As a major advantage of GNN, it is considerably faster than the phase-field model in simulating grain growth. 
Figure \ref{fig:scaling2D} compares the wall times required for phase-field simulation on CPU, GNN on CPU and GNN on GPU to generate a prediction sequence of the same time span.
Xeon Platinum 8160 (2.10 GHz) CPU and NVidia V100 GPU were used in benchmarking.  
Averaged over more than 500 trials, GNN accelerates the 2D predictions by 6 times when run on CPU and 80 times on GPU.   
Even larger acceleration was observed in 3D predictions by GNN, which are faster by 200 times on CPU and 3000 times on GPU than phase-field simulations.
The acceleration achieved by GNN has several origins. 
First, GNN allows for more efficient time stepping. 
While the time step size used by the PDE solvers is limited by the stability of numerical schemes, 
GNN is not subject to the same restriction and could employ larger time steps.
Through testing, we found that the time interval between the two output frames in GNN prediction could be increased to 80 times of the step size in phase-field simulations without affecting the accuracy.
Second, microstructure could be represented on a coarser grid in GNN. 
In phase-field simulations, an interface typically needs to be discretized by at least 5-6 pixels to ensure the numerical accuracy of the difference equations.
In contrast, our test shows that 1-2 pixels is sufficient for GNN, 
which reduces the number of graph nodes required to resolve microstructure by 16 times in 2D and 64 times in 3D compared to phase-field discretization.  
Such spatial coarsening also significantly decreases the computation load.   
In addition, GNN directly benefits from GPU acceleration.
 
\section*{Acceleration with Adaptive Remeshing}
In Ref.~\cite{Pfaff2020GNN} Pfaff and coworkers adopted unstructured triangular cells that are commonly used in finite element methods. Here we adopted square (2D) or cubic (3D) grid with nearest-neighbor edges ($\Box$), as such structured meshes allow for straightforward coding and parallelization. Alternatively, we have tried structured triangles on square grids ($\boxdiag$) but found little difference in accuracy at increased computational costs to treat more edges.

\begin{figure}[thp]
 		\vskip 0.15in
 		\centering
		\includegraphics[width=0.7\columnwidth]{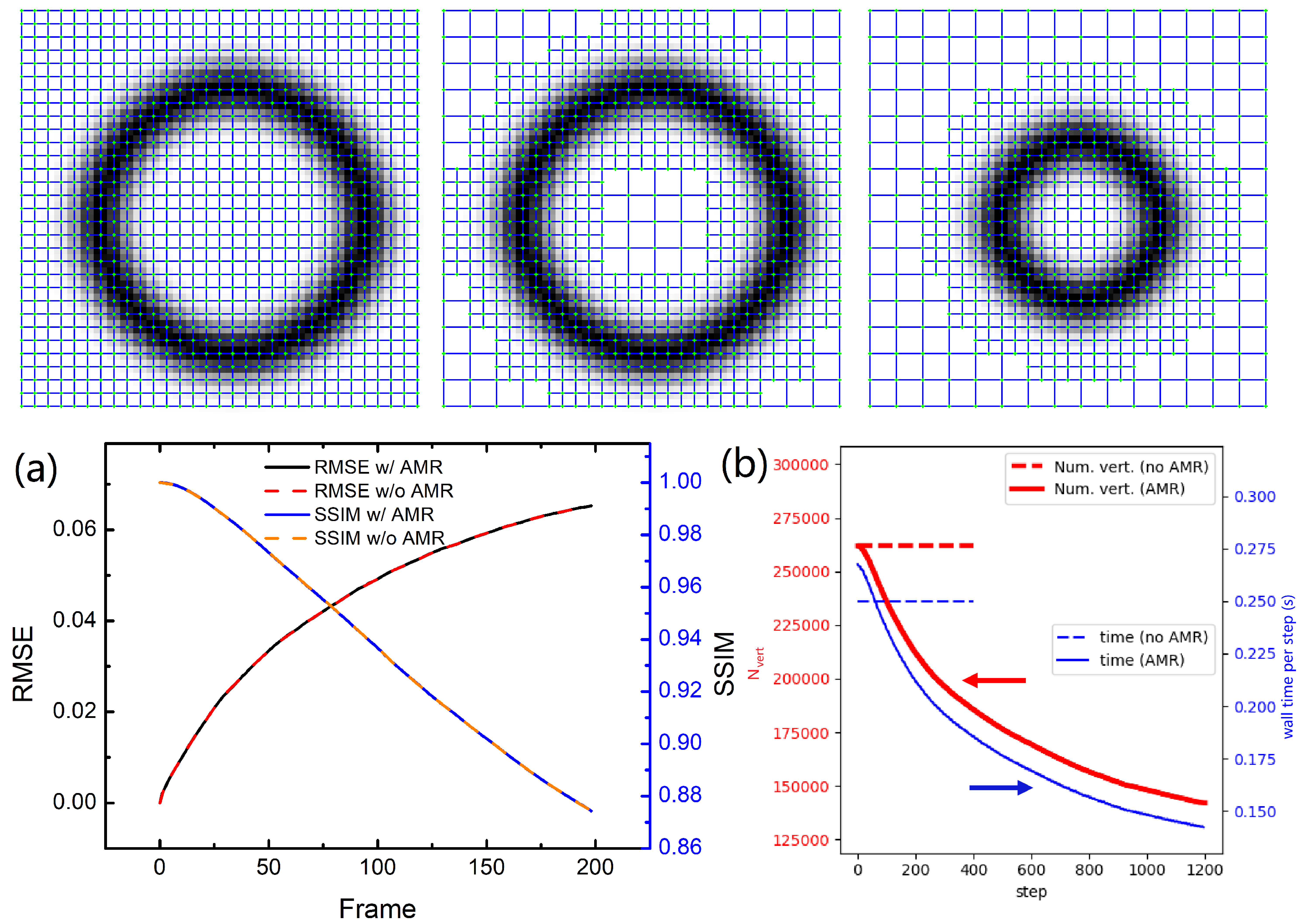}
		\caption{Effects of AMR on GNN simulations on a $512^2$ lattice by remeshing the graph every 5 time steps. (a) Accuracy of the GNN-AMR model compared with the GNN-only model. (b) Average wall time per step (blue solid) and number of vertices $N_\mathrm{vert}$ (thick red solid) versus time step. The number of vertices and average wall time per step for the GNN-only simulation are shown as dashed lines for comparison.}
		\label{fig:AMRtime2D}
\end{figure}

In all previous results, the mesh was kept constant.
However, the initial fine mesh required for small grains becomes excessive for late-stage large grains. To reduce the computational load, we implement adaptive mesh refinement (AMR) within the GNN model. 
The computational details can be found in Section \ref{sec:AMR}.
To test the efficiency gain of AMR, %
a grain configuration defined on a 512$\times$512 fine mesh was used as the initial configuration,
which was evolved by two GNN models trained separately with or without AMR.  
The predictions by the two models are almost identical, which shows that adaptive remeshing has little effect on accuracy (Figure \ref{fig:AMRtime2D}a).  
As shown in Figure \ref{fig:AMRtime2D}b, it takes longer time for the AMR-GNN model (blue curve) to make predictions than the GNN-only model (blue dashed line) at the early stage due to the computational overhead of AMR. 
However, the AMR-GNN model becomes increasingly more efficient when the grain structure gradually coarsens and more and more fine grains are replaced by coarser ones, which is demonstrated by the decrease in the number of vertices (thick red curve) in the AMR-GNN prediction. 
A crossover in the average wall time per step between the two models occurs around 80 time steps,
and the AMR-GNN model becomes twice faster after 1200 steps. 
We find that the computation cost of the AMR-GNN model scales linearly with the number of graph nodes (Figure \ref{fig:AMRtime2D}b). 
Similar behavior is also found in 3D AMR-GNN though with a higher remeshing overhead.
This shows that combining AMR with GNN is especially useful for accelerating long-time predictions of microstructure coarsening processes such as grain growth that involves large change in the structural feature size.   
We note that the current AMR implementation is relatively simple and speedup could be further increased by fine-tuning hyperparameters or employing multilevel grids to represent the graph with coarser cells.

\section*{Acceleration with Adaptive Time Stepping}
\begin{figure}[thp]
	\includegraphics[width=0.9\columnwidth]{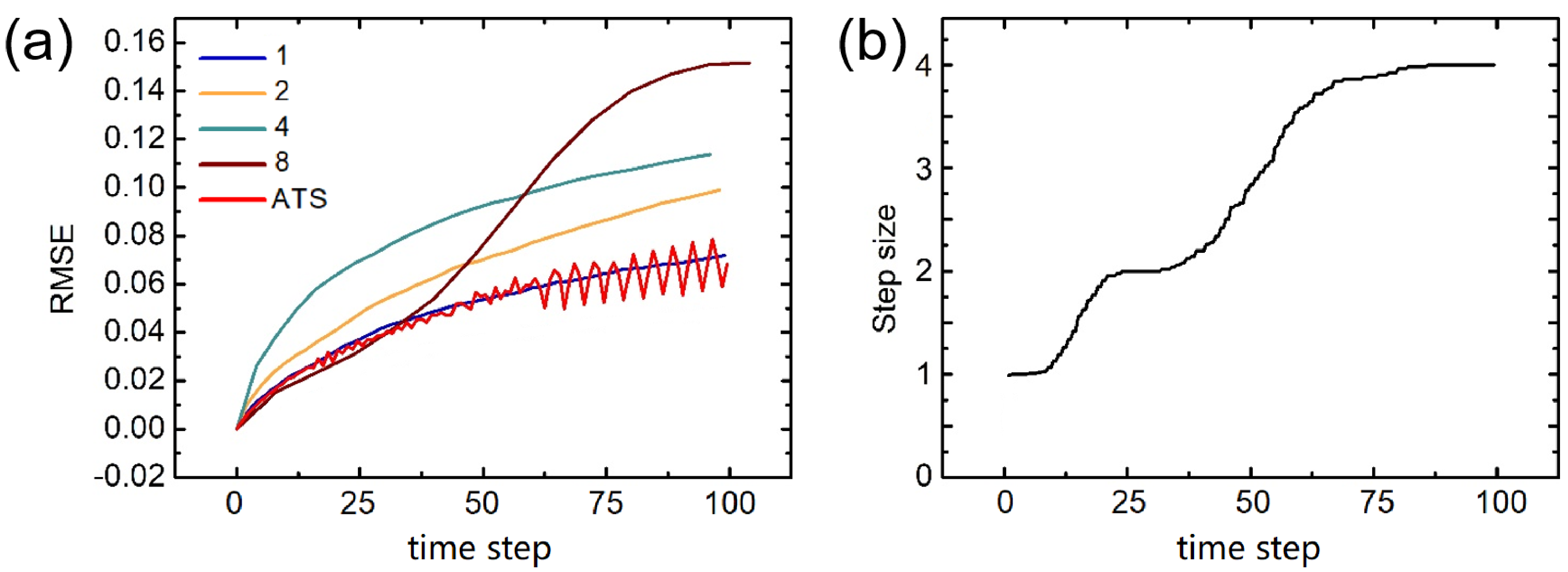}
	\caption{(a) RMSE of the ATS model vs GNN models with different time step size $\Delta t$. (b) Step size averaged over 100 runs vs time steps.}
	\label{fig:adaptive_time_model}
\end{figure}

Like adaptive mesh resolution, adaptive time stepping (ATS) is a commonly employed strategy in accelerating the numerical solution of PDEs. 
Here we applied the ATS approach to further improving GNN's prediction efficiency 
by using increasingly large intervals between output frames.    
When microstructures undergoe coarsening processes such as grain growth and phase separation,
the interface movement becomes progressively slower due to the diminished driving force,  
which naturally allows for larger step size to be used in prediction.
Our approach involves training several GNN models with different time intervals between adjacent frames in the input data. 
The interval for the baseline model is equal to a simulation time of $\Delta t $ = 32. 
Models with larger intervals (2$\Delta t$, 4$\Delta t$) between the adjacent frames were also trained. 
When predicting the evolution sequence, the ATS algorithm switches between models of different $\Delta t$ by using the step-doubling method~\cite{Shampine1985} to dynamically control step size.
Specifically, after every 20 steps, the algorithm compares the predictions using two steps of the current step size versus one step of the doubled step size. 
If their difference measured by RMSE per pixel is smaller than a preset threshold $\lambda$ ($4\times10^{-6}$ was used in this work), the used step size will be doubled in subsequent predictions. 
Figure \ref{fig:adaptive_time_model}a compares the prediction accuracy of this ATS scheme against those of the models with constant step sizes,  
in which the RMSE shown is averaged over 100 runs.
ATS manages to keep the prediction comparable to the model with the smallest step size $\Delta t$. 
On the other hand, the step size (averaged over 100 runs) in the ATS model steadily increases up to 4 times the baseline value over the course of 200 output frames 
(Figure \ref{fig:adaptive_time_model}b), 
which reduces the average simulation time by about 56 percent comparing to the model without ATS.  

\section*{Conclusion}
In summary, we showcase the application of GNN as a surrogate model of computationally intensive PDE-based phase-field methods in simulating materials microstructure evolution by using grain growth as an example. Trained with ground-truth microstructure evolution trajectories from the phase-field model, the GNN model can faithfully predict grain structure evolution in 2D and 3D with both local and statistical accuracy and excellent extrapolation capability in the spatiotemporal domain and configurational space. It substantially improves the quantitative prediction accuracy compared to the 2D CNN model reported in our previous work\cite{Yang2021P} while maintaining a similar computation speed, and provides far superior predictive power in 3D for which CNN qualitatively fails. We also demonstrate that further speedup could be achieved by combining adaptive temporal resolution schemes and especially adaptive mesh refinement, which represents a unique advantage of the graph-based approach with flexible data structures in simulating a wide range of microstructure coarsening phenomena in which the characteristic feature size and time scale continue to increase during the evolution processes.

The present work is limited to the relatively simple case of isotropic grain growth at constant temperature. Future work will attempt to extend the GNN models to the more general situation, where grain growth is anisotropic because of the orientation-dependent grain boundary energy and mobility and also experiences time-dependent and non-uniform temperature during material processing. Such extension could be implemented by incorporating grain orientation and temperature field as additional input channels onto the graph nodes. The training and testing of such models will be reported elsewhere. 

\section*{Computational methods}
\subsection{Phase field method} 
Phase-field simulations are employed to generate the ground truth for grain growth.
	The phase-field method is a powerful computational technique for modeling microstructure evolution in diverse materials systems\cite{Chen2002PFMreview, Moelans2008C, steinbach2009phase}. 
	In a phase-field model, different phases are represented by one or multiple order parameters, and their interfaces are tracked by the level sets of the order parameters. 
	Spatiotemperoal evolution of the microstructure is described by the governing equations of the order parameters derived from thermodynamic and kinetic principles.
	
	In accordance with our previous work \cite{Yang2021P},  isotropic grain growth in polycrystalline structure is simulated by a multi-order-parameter phase-field model\cite{Moelans2008PRB}. 
	In the model, a set of order parameters $\{\eta_1(\boldsymbol{x}), \eta_2(\boldsymbol{x}), ..., \eta_N(\boldsymbol{x})\}$ are used to represent $N$ distinct grain orientations. The free energy of the system is expressed as
	\begin{equation}\label{eq:gg1}
		F = \int \left[ f(\eta_1, \eta_2, ..., \eta_N) + \frac{\nu}{2} \sum^N_{i=1}\left( \nabla\eta_i \right)^2 \right] dV
	\end{equation}
	where the homogeneous free energy density $f$ is given by
	\begin{equation}\label{eq:gg2}
		f = m \left[ \sum^N_{i=1} \left( \frac{\eta^4_i}{4} - \frac{\eta^2_i}{2} \right) + \frac{3}{2} \sum^N_{i=1}\sum^N_{j>i} \eta^2_i \eta^2_j + \frac{1}{4}  \right]
	\end{equation}
	which has $N$ local minima located at $(\eta_1, \eta_2, ..., \eta_N) = (1, 0, ..., 0), (0, 1, ..., 0), ..., (0,0,...,1)$.  
	The evolution of ${\eta_i(\boldsymbol{x})}$ ($i = 1\ldots N$) follows the time-dependent Ginzburg-Landau or Allen-Cahn\cite{allen1972ground,allen1973correction} equation
	\begin{equation}\label{eq:gg3}
		\frac{\partial\eta_i}{\partial t} = -L\frac{\delta F}{\delta \eta_i}
	\end{equation}
	In all the simulations, dimensionless parameters $N=100$, $m= 1$, $\nu=1$ and $L=1$ are used.
	The initial polycrystalline structure is generated by Voronoi tessellation\cite{aurenhammer2000voronoi} with 100 grains in 2d and 350 in 3d. 
	Eq.~\ref{eq:gg3} is solved by the forward Euler finite difference scheme with periodic boundary conditions and grid spacing $\Delta x$ = 1 and time step size $\Delta t$ = 0.2.

\subsection{Data and model setup} 
	Following Ref.~\onlinecite{Yang2021P}, single-channel snapshots of the polycrystalline structure are generated by defining $\phi(\boldsymbol{r}) = \sum^{N}_{i=1} \eta^3_i(\boldsymbol{r})$ as the order parameter $\phi$ so that $\phi$ is close to 0 in the grain boundary region and 1 inside grains. 
The spatial grid of $\phi$ was down sampled spatially  by a factor of $4^2 $ times from $256^2$ to $64^2$ and temporally 80 times, i.e. 80 PDE time steps correspond to one GNN time step. In 3D, the down sampling ratio was $4^3$  spatially and 1000 temporally. Therefore, the data consist of time series of coarse-grained arrays for the field 
$ \phi_t \in \mathbb{R}^{N_x \times N_y (\times N_z) \times N_c}$ 
where $N_{x/y/z}=$64 (2D) or 32 (3D) is the spatial grid size  and $N_c=1$ is the number of channels. The grid size is chosen to be large enough to accommodate sufficient variation in microstructure configurations 
		and also provides adequate resolution to resolve interfaces in microstructure with at least 1-2 pixels.  

The model is implemented in TensorFlow\cite{abadi2016tensorflow} and trained on 2 NVidia V100 or 4 1080-Ti GPUs with single (32-bit) floating-point precision.
	Typical training time is 10 hours, with an initial learning rate of $10^{-3}$ that gradually decays to $10^{-6}$.

In this work, we presume no explicit history dependence, assuming that the next time step is only dependent on the current, not previous steps.
Given  current $\phi_{t}$,  the next frame \(\hat{\phi}_{t+1}\) is predicted by:
\begin{equation} \label{eq:surrogate}
    \hat{\phi}_{t+1} =  \mathbb{G}(\phi_t) + \phi_t
\end{equation}
The time evolution prediction are then performed autoregressively.

\subsection{Message-passing Graph Neural Network} \label{sec:GNN}
We largely adopt the MeshGraphNet of Refs.~\onlinecite{Pfaff2020GNN, Bertin2023-DDGNN}, a type of message-passing graph neural network. 
We used 48 hidden features in 2D and 64 in 3D.

The phase field configuration can be represented by a graph $G = (\mathbb{V},\mathbb{E})$, where $\mathbb{V}=\{i\}$ is the collection of nodes (or vertices, used interchangeably in this work) with nodal positions $\boldsymbol{r}_i$ and field values $\phi(\boldsymbol{r}_i)$, and $\mathbb{E} = \{(i,j)\}$ represents edges or connectivity map. $\mathbb{E}$ is directed: if $(ij) \in \mathbb{E}$, so is $(ji)$. A GNN is an algorithm that naturally operates on such graph-structured data. Specifically, let the input features for each node $i$ be 
\begin{align} \label{eq:node-feature}
V_i = (\phi_i, t_i),
\end{align} 
where $t_i$ is a flag that can be used to specify the type of node $i$. In this work, we the adopted periodic boundary condition such that all nodes are equal (internal): $t_i \equiv 0$.
The input edge features on edge $(ij)$ are 
\begin{align} \label{eq:edge-feature}
E_{ij} =(r_{ij} \equiv |\boldsymbol{r}_{ij}|,  \boldsymbol{r}_{ij} \equiv \boldsymbol{r}_{j}-\boldsymbol{r}_{i})
\end{align} 
Following Ref.~\onlinecite{Pfaff2020GNN, Bertin2023-DDGNN}, our GNN starts with vertex and edge encoders $\text{ENC}^\text{V}$, $\text{ENC}^\text{E}$ transforming concatenated input features into a latent space:
\begin{align} \label{eq:encoder}
    v^{(0)}_i = \text{ENC}^\text{V}(V_i), \ e^{(0)}_{ij} = \text{ENC}^\text{E}(E_{ij}).
\end{align}
Then, $K$ stacked message passing layers $f^{\text{E}(k)}$, $f^{\text{V}(k)}$ ($1 \leq k \leq K$) operate sequentially on the latent edge and vertex features:%
\begin{align}
e^{(k)}_{ij} &= f^{\text{E}(k)}( e^{(k-1)}_{ij}, v^{(k-1)}_i, v^{(k-1)}_j ), \nonumber  \\
v^{(k)}_{i} &= f^{\text{V}(k)}( v^{(k-1)}_i, \sum_j e^{(k)}_{ij}). \label{eq:MP}
\end{align}
Finally, a node decoder $\text{DEC}$ translates the latent node representations $v^{(K)}$ into phase field predictions:
\begin{align} \label{eq:decoder}
    \phi_{t+1}-\phi_t  = \text{DEC}(v^{(K)}_i).
\end{align}
Together, Eqs.~(\ref{eq:node-feature}-\ref{eq:decoder}) constitute a GNN implementation of the surrogate model in Eq.~\eqref{eq:surrogate}.
All neural network operators were built from multi-layer perceptrons with two hidden layers and layer normalization \cite{LayerNorm} between them. Skip connections \cite{resnet} of the form $x \leftarrow x + f(x)$ were adopted in the message-passing layers in addition to Eq.~(\ref{eq:MP}). 
The critical message-passing steps in Eq.~\eqref{eq:MP} allow the GNN to transmit and aggregate information along connected edges, and each additional message-passing layer builds up a more complex dependence on vertices farther away.
The number of message-passing layers $K$ required to obtain accurate predictions is then related to a few factors, including the effective instantaneous interaction distance, and the levels of spatial and temporal downs-sampling.

\subsection{Symmetry and data augmentation}
The data processing and augmentation procedures follow our previous work \cite{Yang2021P}. Each dataset intended for training was randomly partitioned into three subsets: training, validation and testing/prediction (e.g.\ at a ratio of 70:15:15). The validation set was used to monitor the convergence during training, while the testing or prediction set was completely withheld from training. The latter may also include customized sequences with different spatial/temporal dimensions and/or initial configurations. 2D data are augmented by performing symmetry operations of the 2D point group $4mm$ on the original configuration.
Similarly, 3D data are augmented by the cubic $O_h$ point group.

\subsection{Adaptive mesh refinement} \label{sec:AMR}
We assume a fine level $N_1 \times \dots \times N_d$ $d$-dimensional grid. A fine configuration $\phi_{\alpha}(r=\{i_1, \dots i_d\})$ can be specified on this grid, where $\alpha=1,\dots C$ is the input channel index, and $1 \leq i_k \leq N_k$. With a coarsening length of $M$, the grid is divided into $\frac{N_1}{M} \times \dots \frac{N_d}{M}$ subdivisions $\mathbb{S}$, each of which contains $(M+1)^d$ fine grid points (including both interior and surface points on the subdivision) and overlaps with its neighbors at their boundary.  $\mathbb{S}$ is split into a fine subset $\mathbb{F}$ for which all the grid points should be tracked (designated as $T(S)$), as well as a coarse subset $\mathbb{C} = \mathbb{S} \setminus \mathbb{F} = \bar{\mathbb{F}}$ for which only the $2^d$ grid points at the corners are tracked. Further, we designate by $E_\mathrm{F}(S)$ the set of all nearest neighbors fine-level connections of length 1 within the $S$, and by $E_\mathrm{C}(S)$ the set of corner-corner connections of length-$M$ in $S$. For example, in 1D, the first subdivision $S$ consists of $\{1, \dots M+1\}$, the tracked points are $T(S) = \{1, \dots M+1\}$ if $S$ is fine or $\{1, M+1\}$ otherwise, and the edges associated with $S$ are $E_\mathrm{F}(S) = \{\{1,2\}, \dots \{M,M+1\}\} $, $E_\mathrm{C}(S) = \{\{1,M+1\}\} $.
We proposed and implemented several adaptive remeshing algorithms within the TensorFlow framework. The simplest algorithm 1 is:
\begin{itemize}
    \item [1]  Start with the initial configuration $\phi_{\alpha}(r, t=0)$, $\mathbb{F}^{t=0}= \mathbb{S}$, $\mathbb{F}^{t=0}=\emptyset$, and a coarsening tolerance $\delta$.
    \item [2] At time step $t=i$, for each subdivision $S $, compute the amount of fluctuation $\Delta_{S} = \left\lVert  \max_{r \in T(S)} \phi_{r, \alpha} - \min_{r \in T(S)} \phi_{r, \alpha} \right\rVert $, where the norm is taken over the channel dimension.
    \item [3] The updated fine subset is $\mathbb{F}^{i}= \{S |S \in \mathbb{S}, \Delta_S > \delta \} $.
    \item [4] The updated coarse subset is  $\mathbb{C}^{i} = \bar{\mathbb{F}^{i}}$.
    \item [5] For each $S \in \mathbb{F}^{i} \setminus \mathbb{F}^{i-1}$,  interpolate $\phi$ on all $(M+1)^d$ grid points based on the value of $\phi$ on the corners. In this work, we adopted bilinear (trilinear) interpolation in 2(3)-D.
    \item [6] The update nodes are $\bigcup _{S \in \mathbb{S}} T(S)$, or all tracked grid points.
    \item [7] The updated fine edges are $E_\mathrm{F} = \bigcup _{S \in \mathbb{F}} E_\mathrm{F}(S)$, and the coarse edges are $E_\mathrm{C} = \left(\bigcup _{S \in \mathbb{C}} E_\mathrm{C}(S) \right) \setminus E_\mathrm{F}$, i.e.\ those not already connected by length-1 fine edges.
    \item [8] The combined undirected edges are $E_\mathrm{U} = E_\mathrm{F} \bigcup E_\mathrm{C}$, and the directed edges, which are needed in our GNN architecture, are $E_\mathrm{D} = E_\mathrm{U} \bigcup \{\{r_2,r_1\} | \{r_1,r_2\} \in E_\mathrm{U} \}$.
    \item [9] Evolve $\phi$ on the updated graph (nodes and edges), and loop back to step 2.
\end{itemize}
All the steps in the above algorithm can be implemented with efficient parallelizable operators.
Additionally, generalization to the periodic boundary condition is straightforward and will not be detailed here.

However, as we are dealing with sharp grain boundaries, the value of the field $\phi$ may change relatively abruptly between adjacent time steps. As a result, the values on a coarse subdivision may significantly (rather than mildly) exceed the threshold, leading to large potential errors. To remedy this problem, we propose an alternative algorithm 2, whereby we replace step 3 of algorithm 1 with
\begin{itemize}
    \item [3 (v2)] $\mathbb{F'}= \{S |S \in \mathbb{S}, \Delta_S > \delta \} $. The updated fine subset is $\mathbb{F}^{i} = \mathbb{F'} \bigcup \{S | F \in \mathbb{F'}; S, F \ \mathrm{neighbor}  \} $. In this work, $S$ and $F$ are considered neighbors if they share at least a corner.
\end{itemize}
Algorithm 2 adds a buffer layer of fine divisions around those already flagged as fine in algorithm 1. This increase the computational load somewhat but leads to better prediction accuracy.

\subsection{Data analysis}
	The data analysis procedures also follow our previous work \cite{Yang2021P}. RMSE and SSIM are used in pixel-wise comparison between ground truth and predictions.
	RMSE is defined as 
	\begin{equation}
		\text{RMSE} = \sqrt{\sum_{i=1}^{N_x}\sum_{j=1}^{N_y}\frac{(p_\text{g}(i,j) - p_\text{p}(i,j))^2}{N_x N_y}}
	\end{equation}
	where $p_\text{g}(i,j)$ and $p_\text{p}(i,j)$ are the pixel values of ground truth and predictions, respectively.
	SSIM\cite{wang2004image} is defined as 
	\begin{equation}
		\text{SSIM} = \frac{(2\overline{p}_\text{g} \overline{p}_\text{p} + c_1) (2\sigma_\text{gp}+c_2)}{(\overline{p}_\text{g}^2 + \overline{p}_\text{p}^2 + c_1)(\sigma_\text{g}^2 +\sigma_\text{p}^2 + c_2 )}
	\end{equation}
	where $\overline{p}_\text{k}$ and $\sigma_\text{k}$ ($\text{k} = \text{g, p}$) are the average pixel value and variance of ground truth or predictions, respectively, and $\sigma_\text{gp}$ is their covariance. $c_1$ and $c_2$ are small constants and chosen to be $c_1= (0.01L)^2$ and $c_2=(0.03L)^2$, where $L$ is the range of pixel values.
	The Euclidean distance between the distributions of quantity $q$ from RNN predictions and ground truth is defined as 
	\begin{equation}
		d = \sqrt{\sum_{i=1}^n (q_\text{g}^i - q_\text{p}^i)^2}
	\end{equation}
	where $n$ is the number of bins within the interval between the minimum and maximum of $q$, and $q^i_\text{g}$ and $q^i_\text{p}$ are normalized counts in the $i$-th bin of the ground truth and predictions, respectively. $n=20$ is used for all the calculations. 
	
\section*{Acknowledgments}
	S.F. is supported by Department of Energy, Office of Basic Energy Sciences under project number DE-SC0019111.
    A.L.H. acknowledges support from Air Force Office of Scientific Research through Grant No. FA9550-21-1-0460.
    M.T. acknowledges support from the National Science Foundation under project number CMMI-1929949.
B.S.\ acknowledge support by the Laboratory Directed Research and Development (LDRD) program (22-ERD-016) at Lawrence Livermore National Laboratory (LLNL). FZ was supported by the Critical Materials Institute, an Energy Innovation Hub funded by the U.S. Department of Energy, Office of Energy Efficiency and Renewable Energy, and Advanced Manufacturing Office.
The work of B.S.\ and F.Z.\ was performed under the auspices of the U.S. Department of Energy by LLNL under Contract DE-AC52-07NA27344. 
	Phase-field simulations were performed on supercomputers at the Texas Advanced Computing Center (TACC) at The University of Texas.
	GNN training and testing were performed on supercomputers at LLNL and TACC, including allocations from LLNL Institutional Computing Grand Challenge program.

\section*{Author Contributions}
F.Z. conceived the project and implemented the GNN and AMR scheme. F.Z. and M.T. supervised the project. S.F performed phase-field simulations and GNN training and testing and implemented the ATS scheme. A.L.H. performed grain topological evolution analysis. S.F., A.L.H., M.T., B.S. and F.Z. analyzed and discussed the results. S.F., M.T. and F.Z. wrote the manuscript with inputs from other authors.

\section*{Declaration of Interests}
The authors declare no competing interests.

\bibliography{paper}

\end{document}